\documentclass[fleqn,twoside]{article}
\usepackage{espcrc2}

\usepackage{graphicx}
\usepackage[figuresright]{rotating}

\title{CDF Forward Detectors and diffractive structure functions 
\\ at the Fermilab Tevatron
}

\author{Michele Gallinaro~\thanks{for the CDF collaboration}
\address{The Rockefeller University\\ 
	1230 York Avenue, Box 188 \\ New York, NY 10021, USA}
 }

\begin{document}
\begin{abstract}

The CDF Forward Detector upgrade project was designed to enhance the capabilities for diffractive physics at the Tevatron.
It consists of a Roman Pot spectrometer to detect leading antiprotons, a set of counters near and around the beam-pipe to reject the non-diffractive event 
contamination to the data sample, 
and two Miniplug calorimeters to measure the event energy flow in the very forward rapidity region.
In the novel design of the Miniplugs, a lead/liquid-scintillator is read out by wave-length shifting fibers
arranged in a flexible tower geometry and relatively short depth allows calorimetric tracking.
Performance of the Forward Detectors during the first two years of operation in Run~II with colliding proton-antiproton beams at $\sqrt{s}$=1.96~TeV,
as well as the first results obtained, are discussed.
A measurement of the antiproton momentum loss using the Forward Detectors is also presented.

\vspace{1pc}
\end{abstract}


\maketitle

\section{INTRODUCTION}

Diffractive processes at the Tevatron have been studied by tagging events with either a rapidity gap or a leading nucleon~\cite{hcp}.
In order to detect such events, forward regions in pseudorapidity~\cite{rapidity} are extremely important.
At the Fermilab Tevatron collider, proton-antiproton collisions provide the ground to study diffractive processes at high energies.
Two experiments, CDF and D\O, collected data in the 1990's at an energy of $\sqrt{s}$=~1.8~TeV 
and continue to do so in the first decade of this century with new upgraded detectors 
during the second phase of data-taking at $\sqrt{s}$=~1.96~TeV.
These two periods are usually referred to as Run~I and Run~II, respectively.
Diffractive physics topics to be addressed in Run~II include 
studies of soft and hard diffraction, forward jet production, and exclusive production of dijet, low-mass and heavy flavor states. 

The Run~II physics program at the Tevatron Collider started in November 2001.
Both the CDF and the D\O~experiments underwent major upgrades to improve their detector capabilities.
Among these, the Forward Detector~\cite{fd} upgrade project at CDF will enhance 
the sensitivity for hard diffraction and very forward physics during Run~II.

The Forward Detectors include the Roman Pot
fiber tracker spectrometer (RPS) to detect leading antiprotons,
a set of Beam Shower Counters (BSCs)
installed around the beam-pipe at three (four) locations along the $p$($\overline p$)
direction to tag rapidity gaps at 5.5~$<|\eta |<$~7.5, 
and two forward MiniPlug (MP) calorimeters covering the pseudorapidity
region 3.5~$<|\eta |<$~5.1.
All these detectors have been installed and are now collecting data, fully integrated with 
the remainder of the CDF detector.

\begin{figure}[htb]
\includegraphics*[width=\hsize]{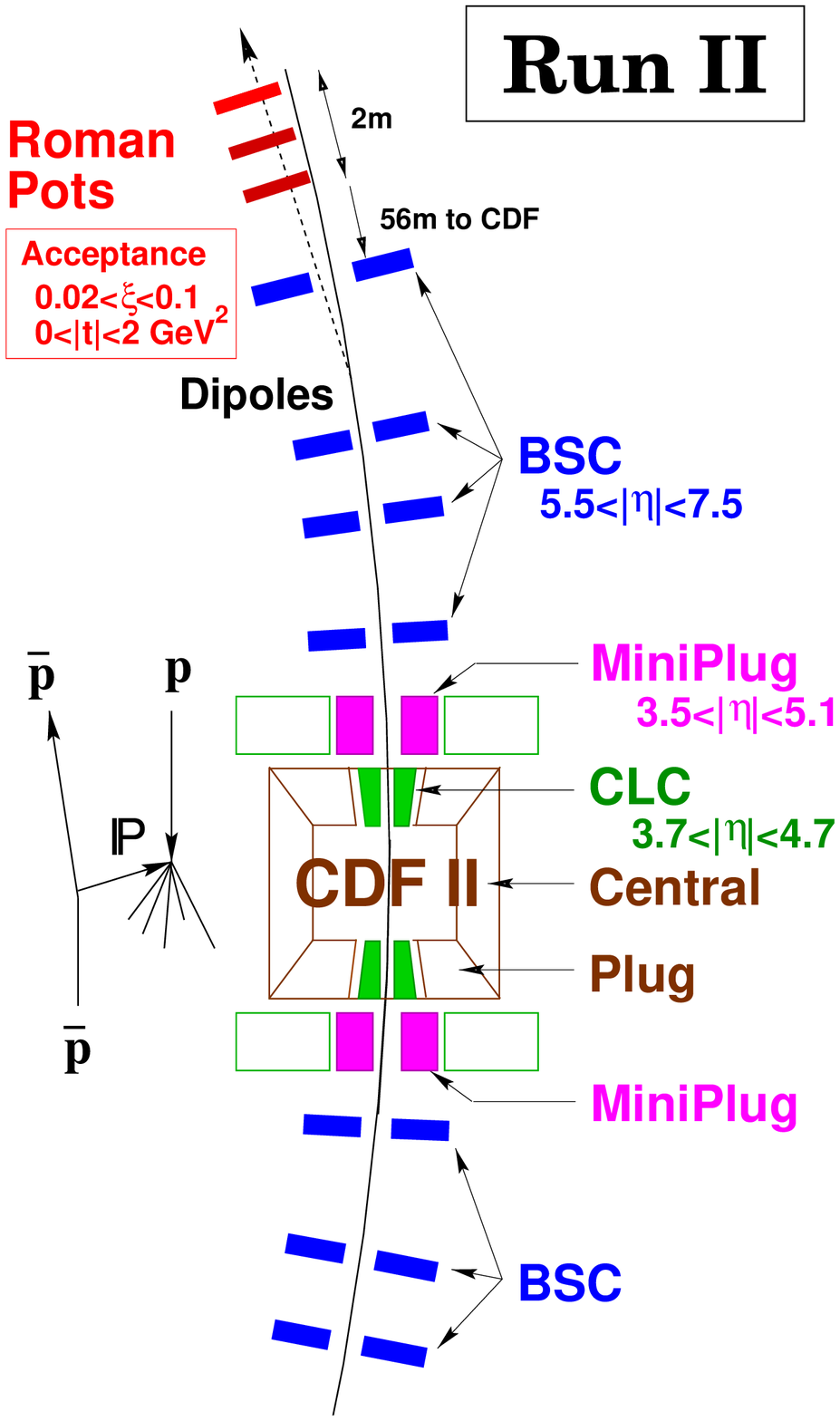}
\caption{\label{fig:fd_top} Forward detector location along the beam-pipe (not to scale).}
\end{figure}

\section{ROMAN POT FIBER TRACKER}

\begin{figure}[htb]
\includegraphics*[width=\hsize]{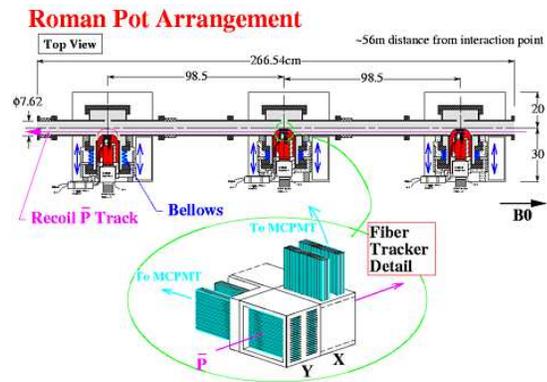}
\caption{\label{fig:rps_detector} Schematic view of the Roman Pot fiber tracker detector.}
\end{figure}

The pomeron structure can be determined by using the kinematic variables in
diffractive dijet events.
During Run~I, the momentum fraction of the $\overline p$ carried by the pomeron 
was determined using the RPS to measure the momentum of the leading anti-proton.
The RPS is a fiber detector spectrometer with a 2-m lever arm
located approximately 56~m from the interaction point (IP), downstream 
of the anti-protons (Fig.~\ref{fig:fd_top}).
It consists of three stations, approximately 1~m apart from each
other (Fig.~\ref{fig:rps_detector}).
Each station comprises one trigger counter and one 80-channel scintillation fiber detector
viewed by a Hamamatsu H5828 80-channel photo-multiplier tube~(PMT).
A coincidence of the three trigger counters selects events with an outgoing
anti-proton at the RPS location.

\begin{figure}[htb]
\includegraphics*[width=\hsize]{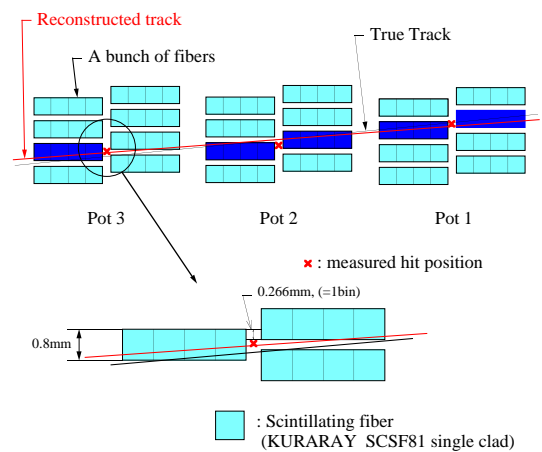}
\caption{\label{fig:rp_track} A schematic drawing of a track traversing the RPS fiber tracker (not to scale). 
Only four out of twenty fiber bunches are shown here for each plane.}
\end{figure}

The fiber detector reads X (40 channels) and Y (40 channels) 
coordinates to identify the position of the tracks with a resolution 
of approximately $100~\mu$m.
The fibers are layered in {\it bunches} of four (Fig.~\ref{fig:rp_track}) in order to increase the 
scintillating material traversed by the incoming particle.
The fibers have a transversal section of 0.8~mm$\times$0.8~mm
and are approximately 20~cm long. Each bunch has a total thickness of 3.2~mm and is read out by a single PMT channel.
Each RPS station has four planes for reconstructing X and Y coordinates
(two for X and two for Y).
Both in the X and in the Y directions,
the fibers of the two planes are shifted laterally relative to each other by one half of the fiber thickness for better spatial 
resolution and overlap by 266~$\mu$m (approximately one third of their thickness).
Figure~\ref{fig:rp_track} shows a section of the fiber tracker 
in one direction (either X or Y).

In preparation for Run~II, the RPS was reinstalled as in Run~I.
In contrast, the readout electronics was completely redesigned to take into 
account the shorter (with respect to Run~I) 396-nsec~spacing between bunches.
Moreover, the beam polarity at the RPS location was reversed with respect to that in Run~I,
so that the anti-proton beam travels now closer to the RPS detector.
Typically, during data-taking, the sensitive detectors of the RPS are located at 
approximately 1~cm from the beam.

\section{BEAM SHOWER COUNTERS}

Single diffractive (SD) and double pomeron exchange (DPE) processes are characterized by forward rapidity gaps. 
A rapidity gap tagger at the trigger level would select these
processes and can be accomplished with a set of scintillation counters around 
the beam-pipe at several locations along the $p$ and $\bar{p}$ directions 
covering the forward pseudorapidity region.

The BSCs are used to identify diffractive events with a leading anti-proton
by rejecting non-diffractive (ND) minimum bias events and events due to multiple interactions with at least one overlapping ND event.
These counters are used to reduce ND background at the trigger level and thus make it possible to collect diffractive data at high luminosities.

\begin{figure}[htb]
\includegraphics*[width=\hsize]{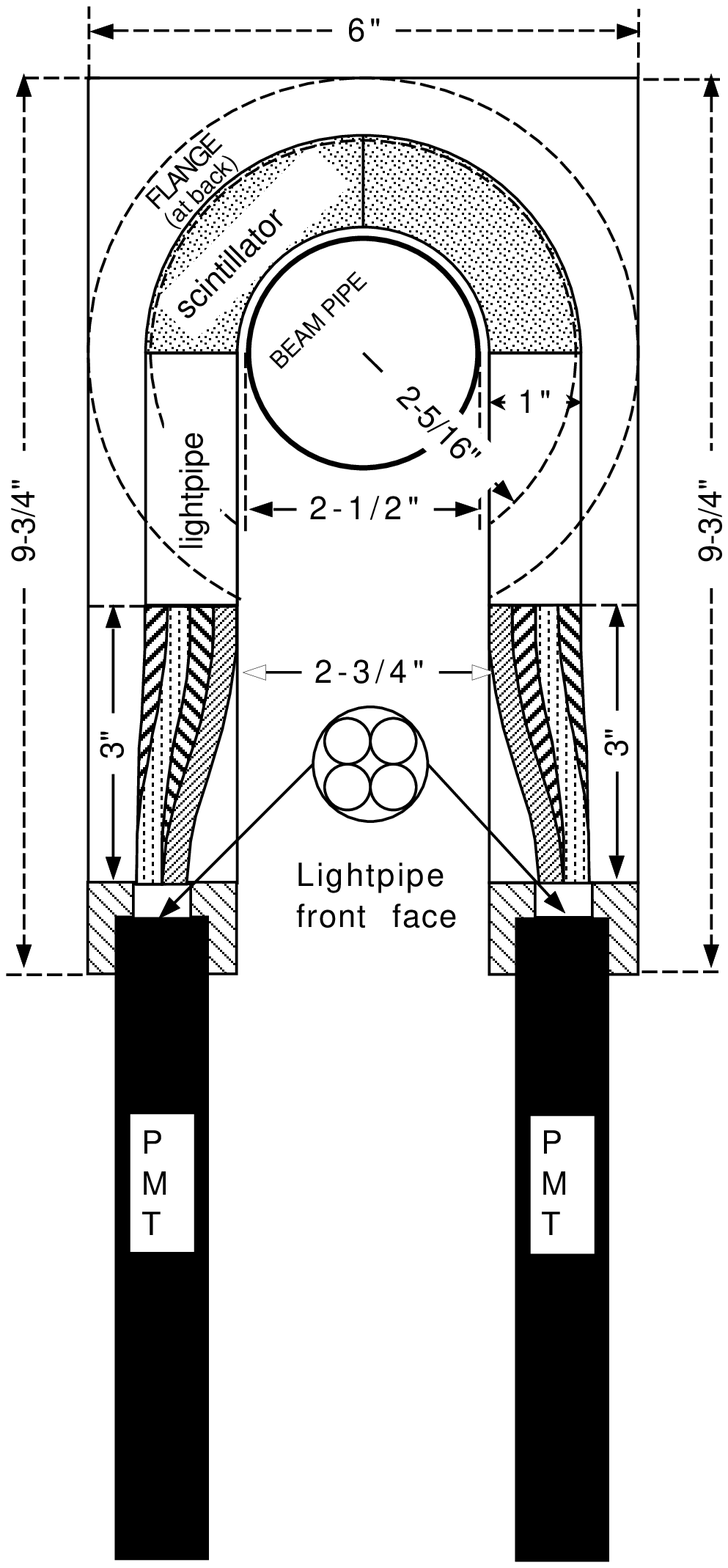}
\caption{\label{fig:bsc1_drawing} Schematic drawing of one half of the beam shower counters BSC-1.}
\end{figure}

The BSCs detect particles traveling in either direction from the IP along 
and near the beam-pipe and cover the pseudorapidity region 5.5~$<|\eta |<$~7.5.
There are four BSC stations on the west side and three on the east side of
the IP (Fig.~\ref{fig:fd_top}).
All stations are located along the beam-pipe, at increasing distances from the IP as one goes from BSC-1 to BSC-4.  
BSC-1, 2 and 3 consist of two stations each, positioned symmetrically with 
respect to the IP, whereas BSC-4 consists of a single station on the west side.
Stations are made of two scintillation counters, except for the BSC-1 
stations, which have four counters (Fig.~\ref{fig:bsc1_drawing}).
Since each counter is connected to its own PMT, the entire system 
consists of 18 signal channels, 10 from the west 
and 8 from the east side.

\begin{figure}[htb]
\includegraphics*[width=\hsize]{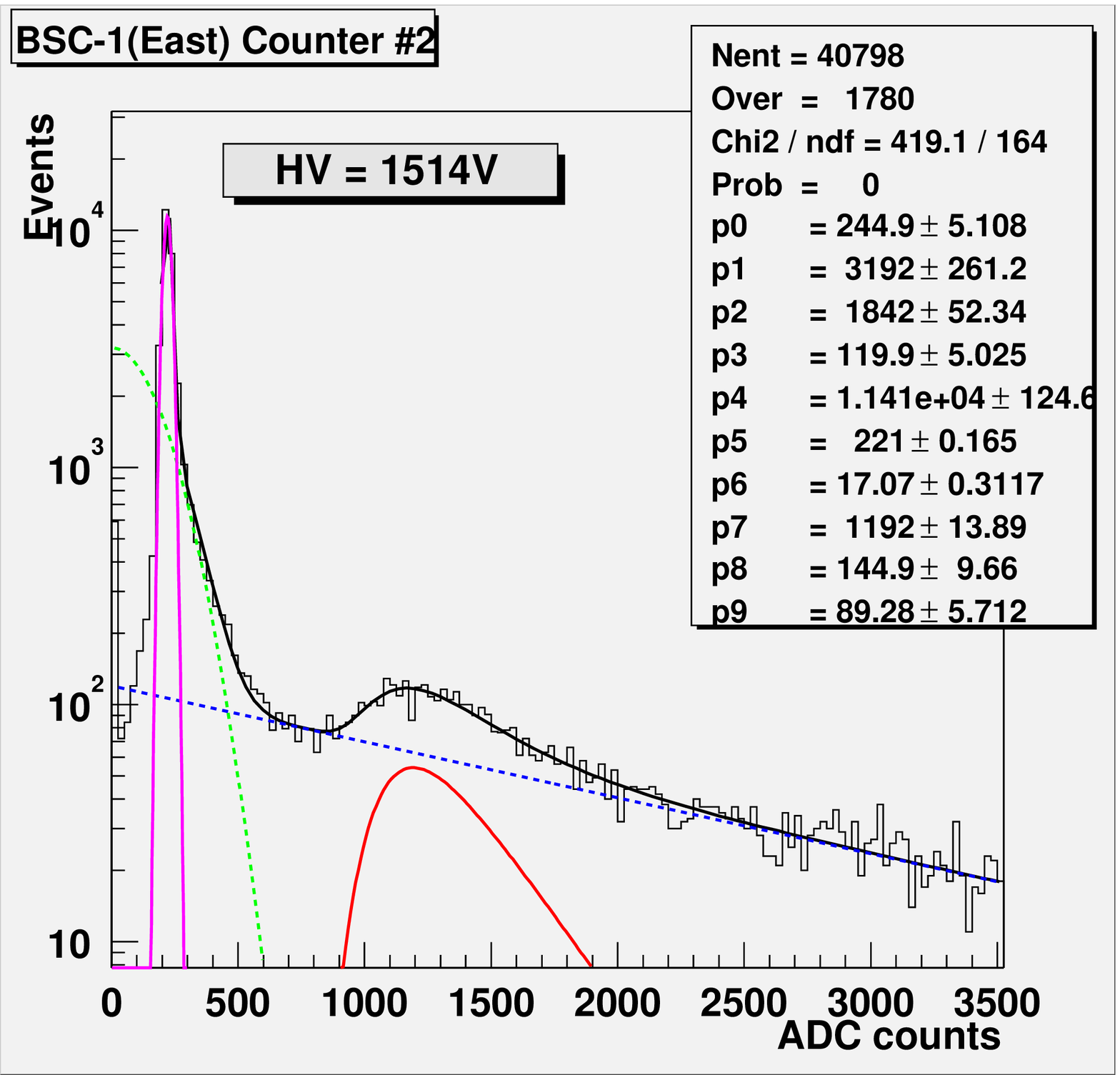}
\caption{\label{fig:bsc1_mip} ADC count distribution for one of the counters in the BSC-1 East station.
A broad MIP peak is visible at a value of about 1200~ADC counts.}
\end{figure}

The distributions of ADC counts from the counters were compared 
in order to adjust the gains.
The shapes of the distributions show good agreement and indicate that 
the gains are well equalized.
Figure~\ref{fig:bsc1_mip} shows the fit to a 
minimum ionizing particle (MIP) peak in 
the ADC distribution for one of the
BSC-1 counters. As for the distributions for the other BSC-1 counters, 
the gains have been adjusted so that the MIP peaks measure 
approximately 1000~ADC counts.

The BSC-1 counters are also used to monitor Tevatron beam losses and collision rates.
These counters cover the pseudo-rapidity region of $5.5<|\eta|<5.9$.
The beam loss rates provide a beam quality index to be used for selecting appropriate beam conditions.
The signals from BSC-1 are referred to as ``out-of-time'' for particles hitting
the counters $\sim 20$~nsec before the interaction time (also due to beam losses),
and as ``in-time'' for particles coming from a collision and hitting
the counters $\sim 20$~nsec after the interaction time (collision rate and diffractive
physics measurements).

\clearpage

\section{MINIPLUG DETECTORS}

\begin{figure}[htb]
\includegraphics*[width=\hsize]{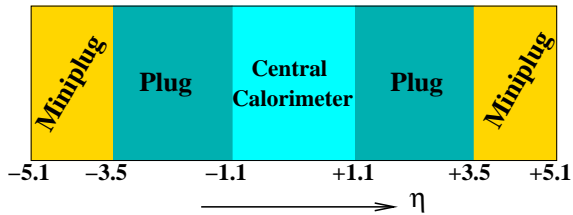}
\caption{\label{fig:cal_rapidity} Rapidity coverage of CDF calorimeters in Run~II: central, plug and Miniplug.}
\end{figure}

The event energy flow in the very forward direction is measured by the MP calorimeters.
The diffractive physics and very forward physics program for Run~II benefits from
two forward MP calorimeters 
designed to measure the energy and lateral position of both
electromagnetic and hadronic showers in the region $3.5<|\eta|<5.1$.
The MPs can detect both charged and neutral particles.
They extend the pseudorapidity
region covered by the Plug calorimeters, which is 1.1~$<|\eta|<$~3.5.
It should be noted that the entire CDF detector, both in the central and in the forward rapidity regions,
is essential to the analysis of diffractive processes.
Figure~\ref{fig:cal_rapidity} shows the calorimetry coverage of CDF in Run~II.

\begin{figure}[htb]
\includegraphics*[width=\hsize]{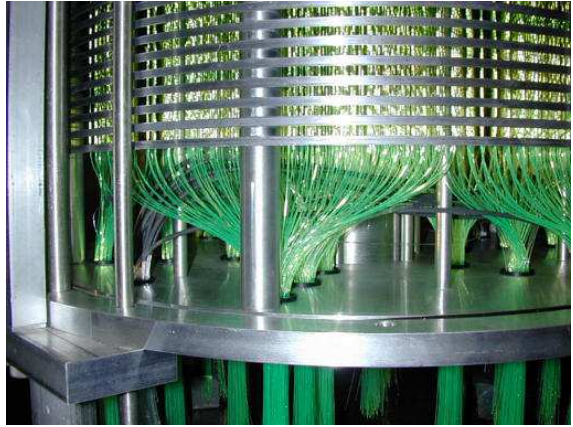}
\caption{\label{fig:mp_fibers} Fiber routing in the Miniplug.}
\end{figure}

The MP and Plug calorimeters can measure the width of the rapidity gap(s)
produced in diffractive processes and will allow extending Run~I studies of
the diffractive structure function to much lower values of the fractions $\xi$,
where $\xi$ is the momentum of the proton carried by the pomeron. 

\begin{figure}[htb]
\includegraphics*[width=\hsize]{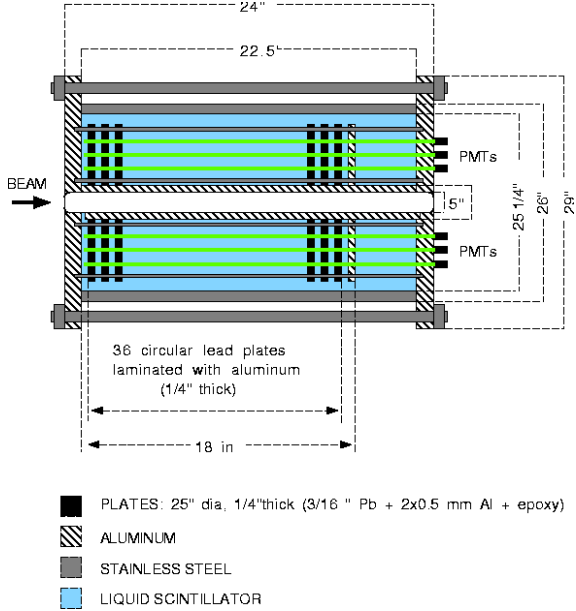}
\vspace*{-1.0cm}
\caption{\label{fig:mp_side}
Side view of a Miniplug (not to scale).}
\end{figure}

The MPs consist of alternating layers of lead plates and liquid scintillator 
read out by wave-length shifting~(WLS) fibers (Fig.~\ref{fig:mp_fibers}).
The WLS fibers are perpendicular to the lead plates and parallel to the 
proton/anti-proton beams, in a geometry where towers are formed by combining
the desired number of fibers
and read out by multi-channel photomultipliers~(MCPMTs).
The 16-channel R5900 MCPMTs have been produced by Hamamatsu with a quartz window 
that significantly improves the radiation hardness.
The MP has a ``towerless'' geometry and no dead regions due to the 
lack of internal mechanical boundaries.
Each MP is housed in a cylindrical steel barrel  $26''$ in diameter and has a 
$5''$--hole concentric with the cylinder axis to accommodate the 
beam-pipe (Fig.~\ref{fig:mp_side}).
The active depth of each MP is 32 radiation lengths and 1.3 interaction lengths.
The ``short'' hadronic depth confines the lateral spread of the 
showers, thereby facilitating the determination of the shower position and particle counting.

\begin{figure}[htb]
\includegraphics*[width=\hsize]{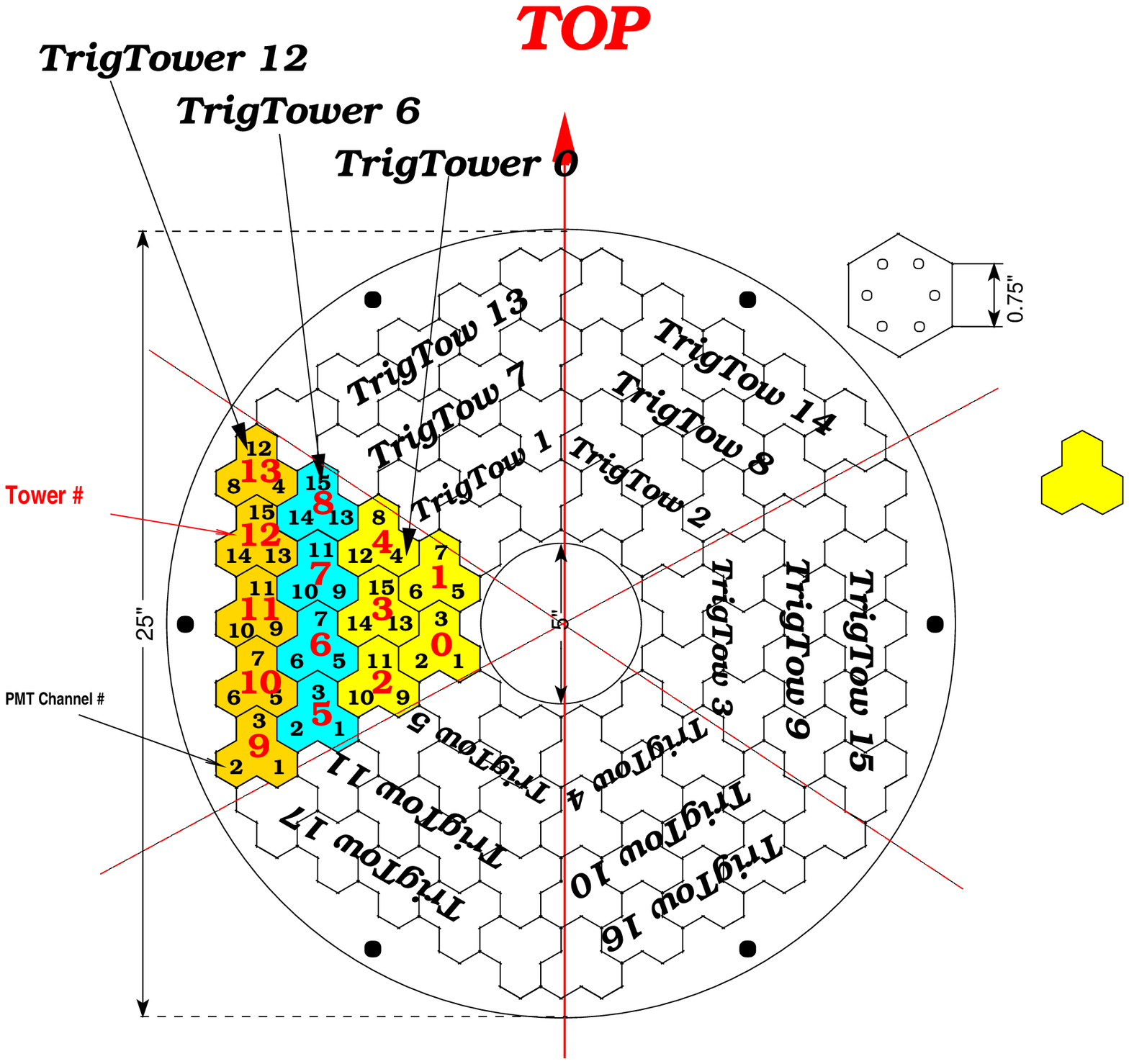}
\vspace*{-1.5cm}
\caption{\label{fig:mp_tower_geometry} 
Tower geometry of the East Miniplug calorimeter (viewed from the interaction point).}
\end{figure}

The design is based on a hexagon geometry. Uniformly distributed over each
plate, holes are conceptually grouped in hexagons with each hexagon comprising six holes.
A WLS fiber is inserted in each hole. The six fibers of one hexagon are grouped
together and viewed by one MCPMT channel.
The MCPMT outputs are added in groups of three to form 84 calorimeter towers in order to
reduce the costs of the readout electronics.
The tower geometry is organized in four concentric circles around the beam-pipe
(Fig.~\ref{fig:mp_tower_geometry}).

The sum of all MCPMT channels is also read out through the last dynode output,
indicated as {\it TrigTower} in Figure~\ref{fig:mp_tower_geometry},
to provide triggering information.
Each MP has a total of 18 trigger towers, arranged in three rings --
the {\it inner}, the {\it middle} and the {\it outer} ring. 
This allows triggering on different pseudorapidity regions, either for events with a 
{\it gap} region or for events with large energy clusters.
An additional clear fiber carries the light from a calibration light-emitting diode (LED) to each MCPMT pixel.
The LED allows a relative gain calibration to equalize the MCPMT gains 
and also allows periodical monitoring of the MCPMT response.

\begin{figure}[htb]
\includegraphics*[width=\hsize]{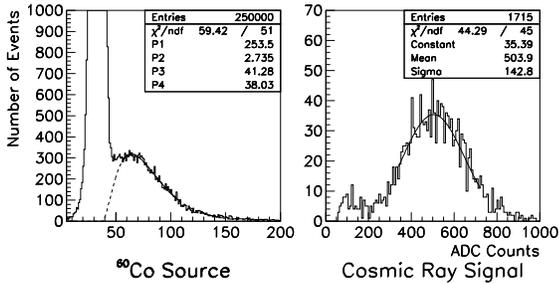}
\vspace*{-1.5cm}
\caption{\label{fig:cosmic_test}
Cosmic ray test of {\it Tower} \#7 of the east Miniplug.
$^{60}$Co source signals; {\it P3} parameter corresponds to the
pulse height of a single photoelectron (left).
Cosmic ray spectrum after an isolation cut fitted to a
Gaussian distribution (right).}
\end{figure}

Cosmic ray muons were used to test one 60$^\circ$-wedge of the East MP.
In this test, the cosmic ray trigger fired on a 2-fold coincidence 
of scintillation 
counter paddles located on top and at the bottom of the MP vessel,
placed with the towers pointing upward.
The outputs from Towers~\#5,~6,~7 and~8 and from Trigger Towers~\#0,~6 and~12 
(Fig.~\ref{fig:mp_tower_geometry}) were read out.
An energy isolation cut selected only those muons which went
through the entire length of the central Trigger Tower (\#6) and vetoed 
on the signals from the neighboring Trigger Towers (\#0 and 12).
To calibrate the response to MIPs in photoelectrons, the single photoelectron response for Tower \#7 was measured using 
a randomly gated signal from a $^{60}$Co source 
(Fig.~\ref{fig:cosmic_test}).
The single tower response to a MIP was found to be
approximately 120 photoelectrons, exceeding design specifications.

\begin{figure}[htb]
\includegraphics*[width=\hsize]{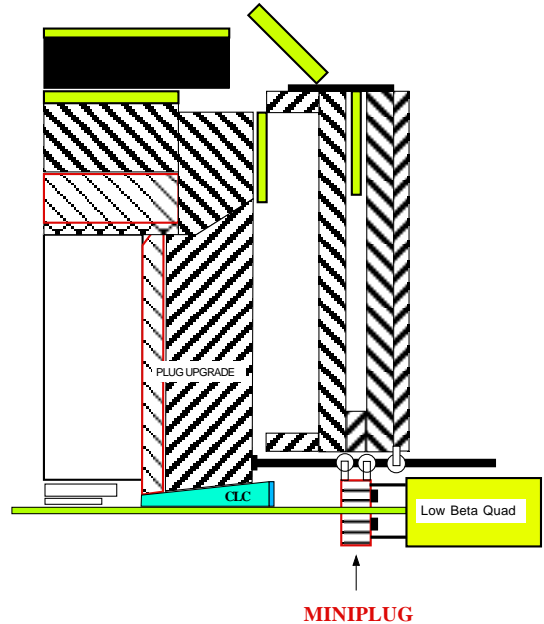}
\vspace*{-1.4cm}
\caption{\label{fig:mp_at_cdf}
Schematic drawing of a quarter view of the CDF detector showing a
Miniplug calorimeter installed inside the toroids (not to scale).}
\end{figure}

The MPs have been installed along the beam-pipe within the hole of
the muon toroids at a distance of 5.8~m from the center of the CDF detector
(Fig.~\ref{fig:mp_at_cdf}). 
The MP detectors are now fully instrumented and have been collecting data since June 2002.

\section{MINIPLUG CALIBRATION}

\begin{figure}[htb]
\includegraphics*[width=\hsize]{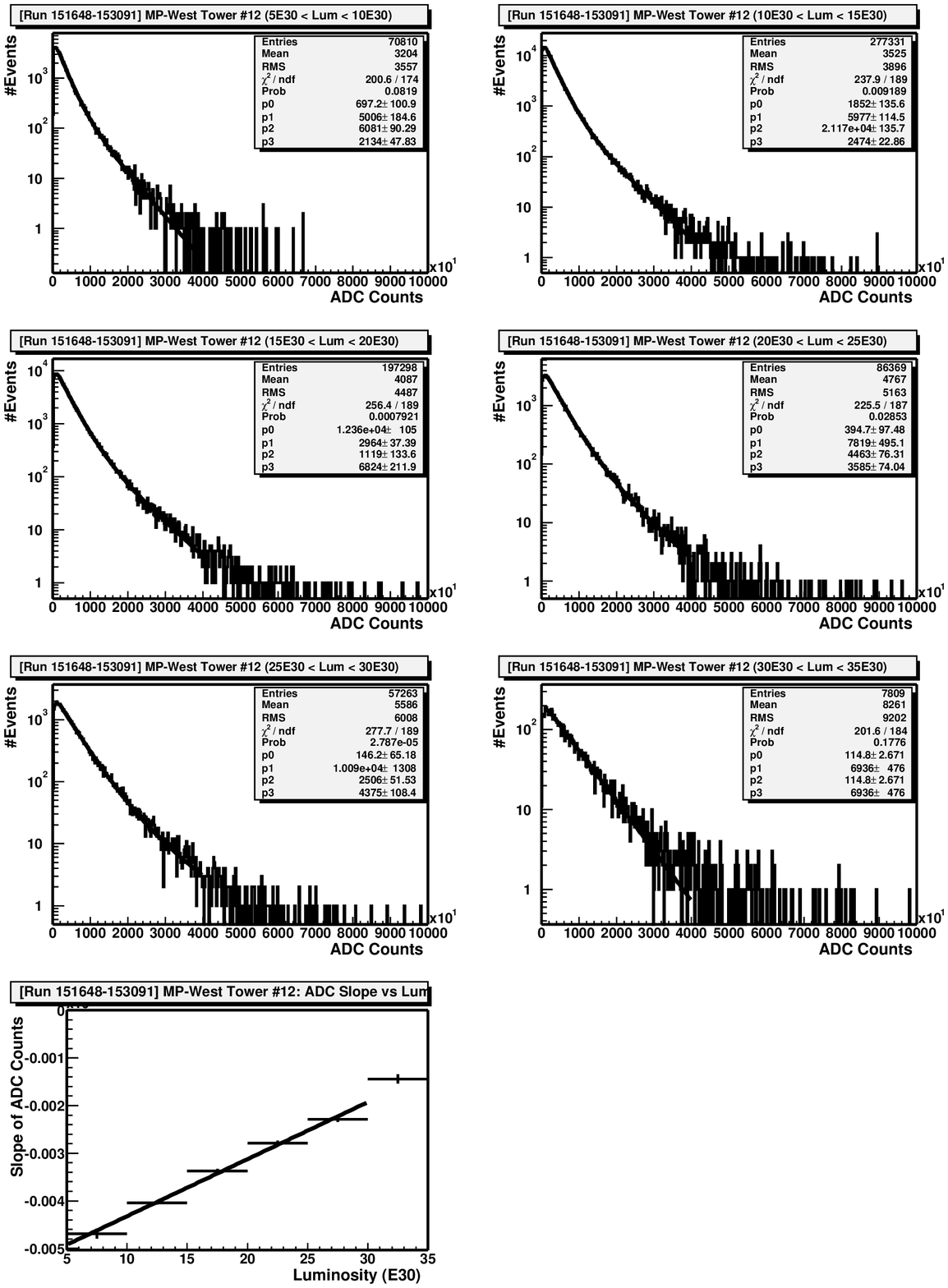}
\vspace*{-1.5cm}\caption{\label{fig:mp_adc} The ADC count distribution of the data can be fitted well with
a falling exponential curve. The curve is a double exponential fit to the distribution.
The ADC distribution of the West Miniplug tower \#12 in the outer layer is shown.}
\end{figure}

\begin{figure}[htb]
\includegraphics*[width=\hsize]{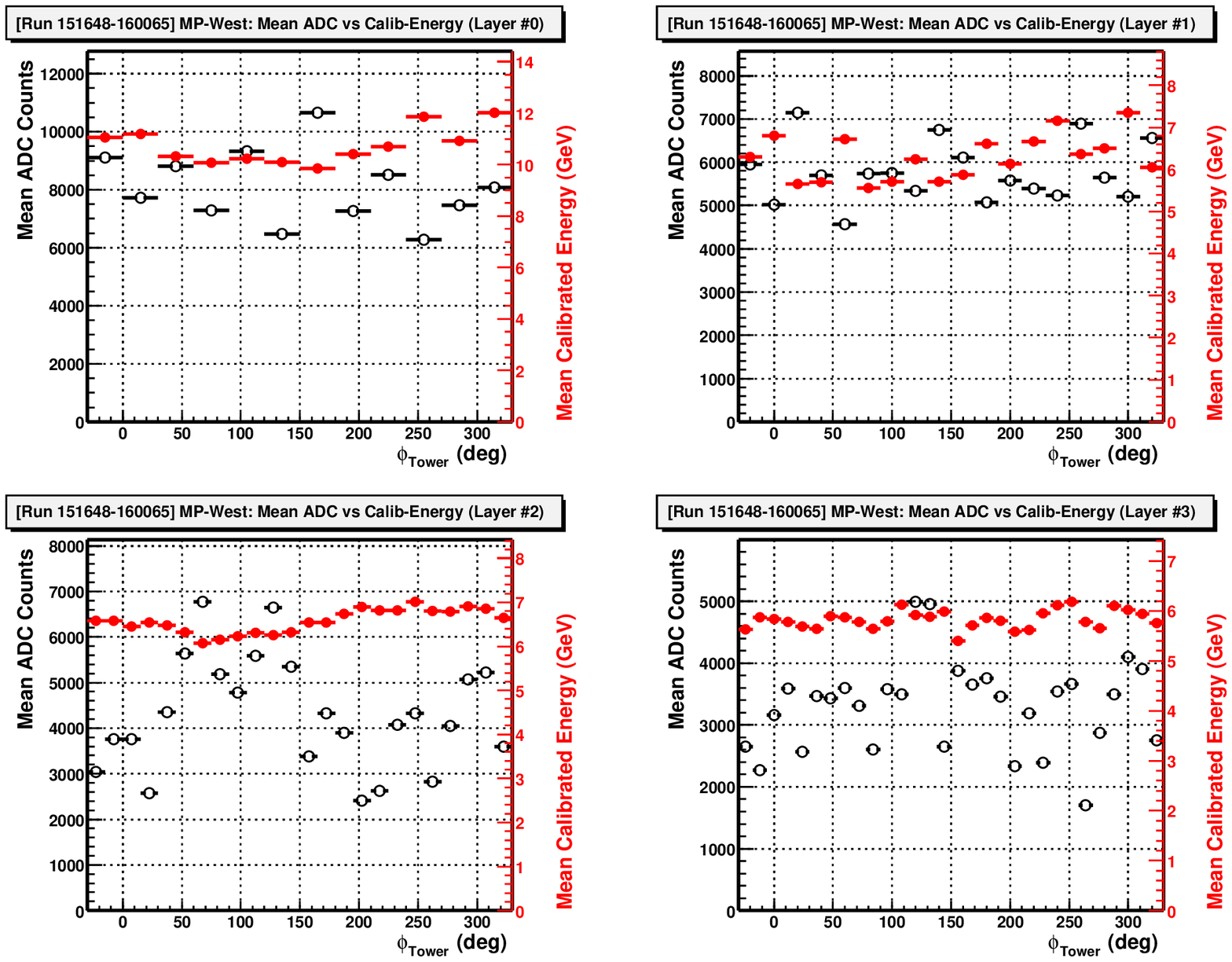}
\caption{\label{fig:mp_encal} Mean values of raw ADC counts (open circles) and calibrated energy (full circles).}
\end{figure}

The first data from Run~II have been used to calibrate and commission the MPs.
Although a precise energy calibration of the MP is not crucial to the
understanding of diffractive processes, an attempt was made 
to estimate the energy scale of jets and particles.
To this end, a Monte Carlo simulation
was used to calibrate the pseudorapidity dependence of the particles' energies and thereby 
the tower-by-tower relative response.
For each tower, the ADC count distribution of the data can be fitted 
well with a falling exponential curve, as shown in Figure~\ref{fig:mp_adc},
and is compared with a Monte Carlo simulation for a sample of minimum bias events.
The slopes are first equalized separately in each ring and then adjusted to the 
slopes predicted from Monte Carlo for different pseudorapidity regions.
Figure~\ref{fig:mp_encal} shows the mean raw ADC counts values of all MP towers (open circles) in one of the four rapidity rings.
The horizontal axis covers the entire range in azimuthal angle $\phi$.
MP tower response becomes flat within $\pm 20\%$ after energy calibration (full circles).

\begin{figure}[htb]
\includegraphics*[width=\hsize]{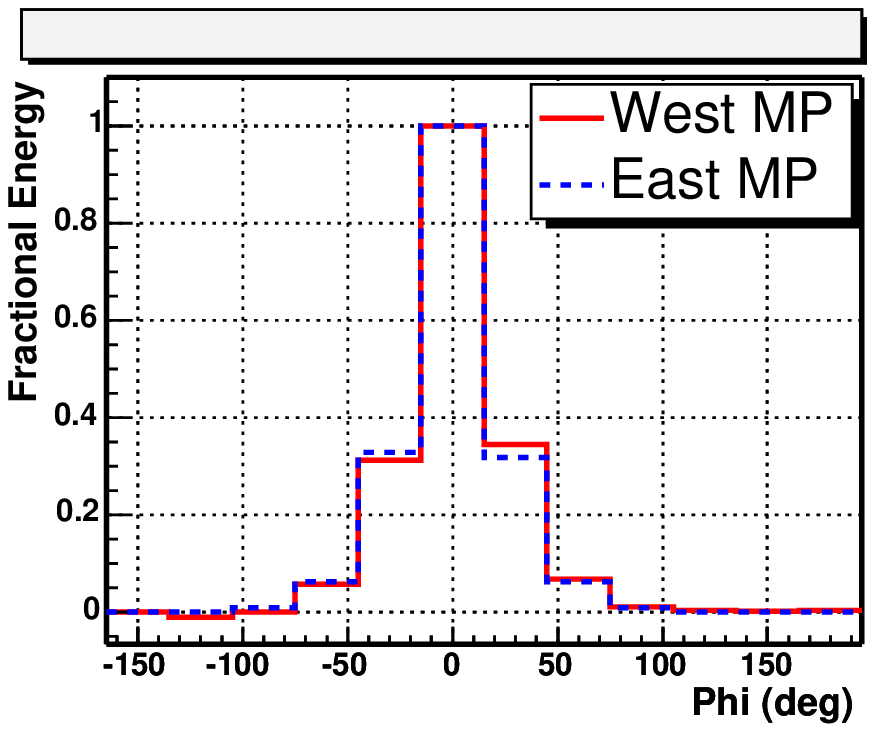}
\vspace*{-1.5cm}
\caption{\label{fig:mp_seed}
Fractional energy of MP towers in the inner $\eta$ layer with respect to the energy of the seed tower, which is always set at $\phi=0$. 
Solid (dashed) histograms show the distributions for the West (East) Miniplug.}
\end{figure}

\begin{figure}[htb]
\includegraphics*[width=\hsize]{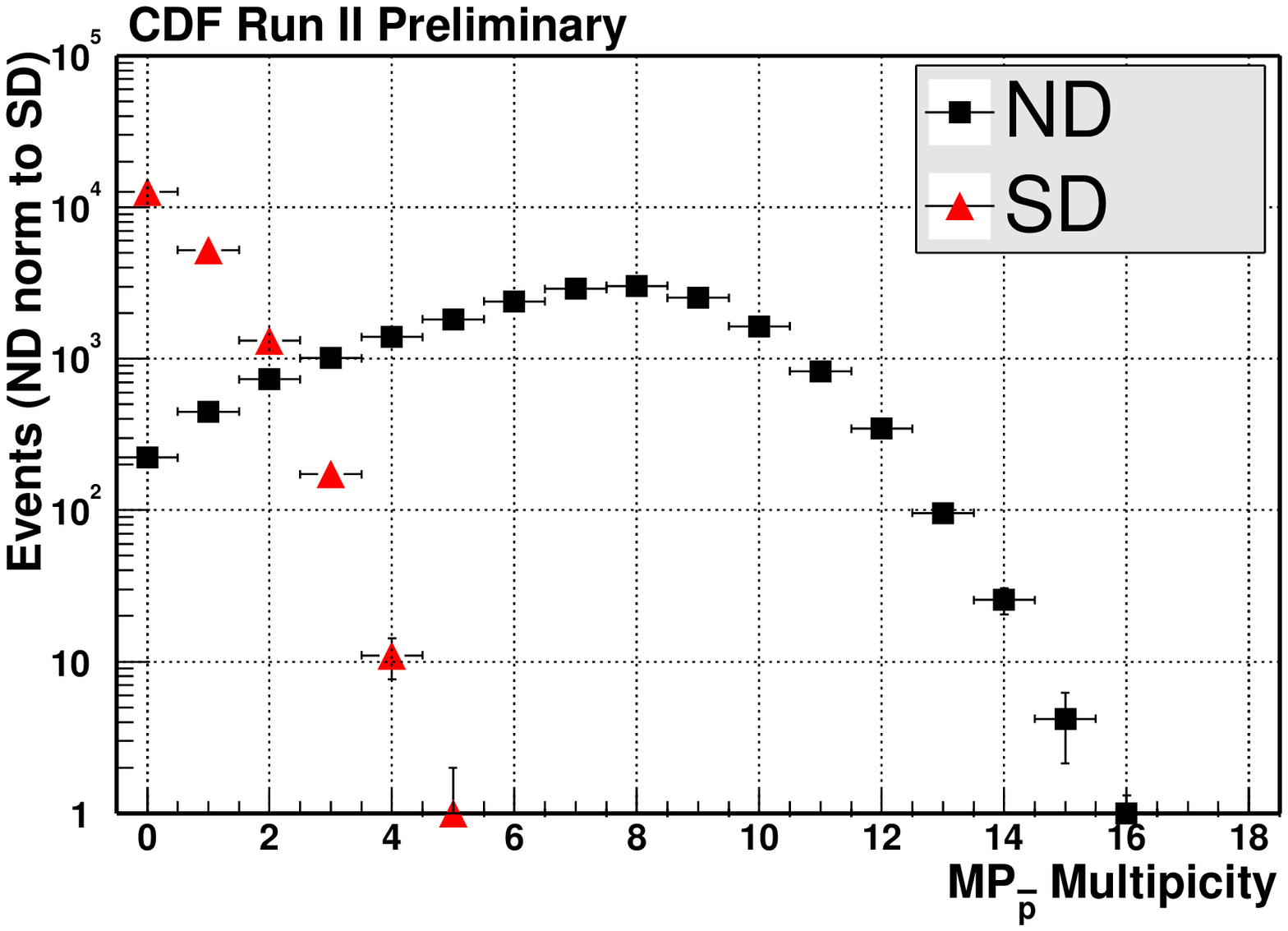}
\caption{\label{fig:mpw_mult}
Particle multiplicity in West MP for diffractive (SD) and non-diffractive (ND) events.}
\end{figure}

Due to the flexible tower geometry in the MP, the position of the particle initiating the shower is found as the center of the towers hit.
The initiated shower also spreads to the neighboring towers (Fig.~\ref{fig:mp_seed}).
In fact, about 25\% of the energy of the shower is deposited in the seed tower, while the remainder is deposited in the surrounding towers.
Particle multiplicity in the MP is calculated by counting the number of ``peaks'' above detector noise,
using ``seed'' towers with a minimum transverse energy of $E_T>200$~MeV. 
The West MP multiplicity is depicted in Figure~\ref{fig:mpw_mult}.
The SD events have smaller multiplicity in comparison with ND events.
Although the energy measured in the MP is not determined with the same accuracy as in the central 
calorimeter, the $\xi$ measurement and the selection of diffractive events are not appreciably affected.

\begin{figure}[htb]
\includegraphics*[width=\hsize]{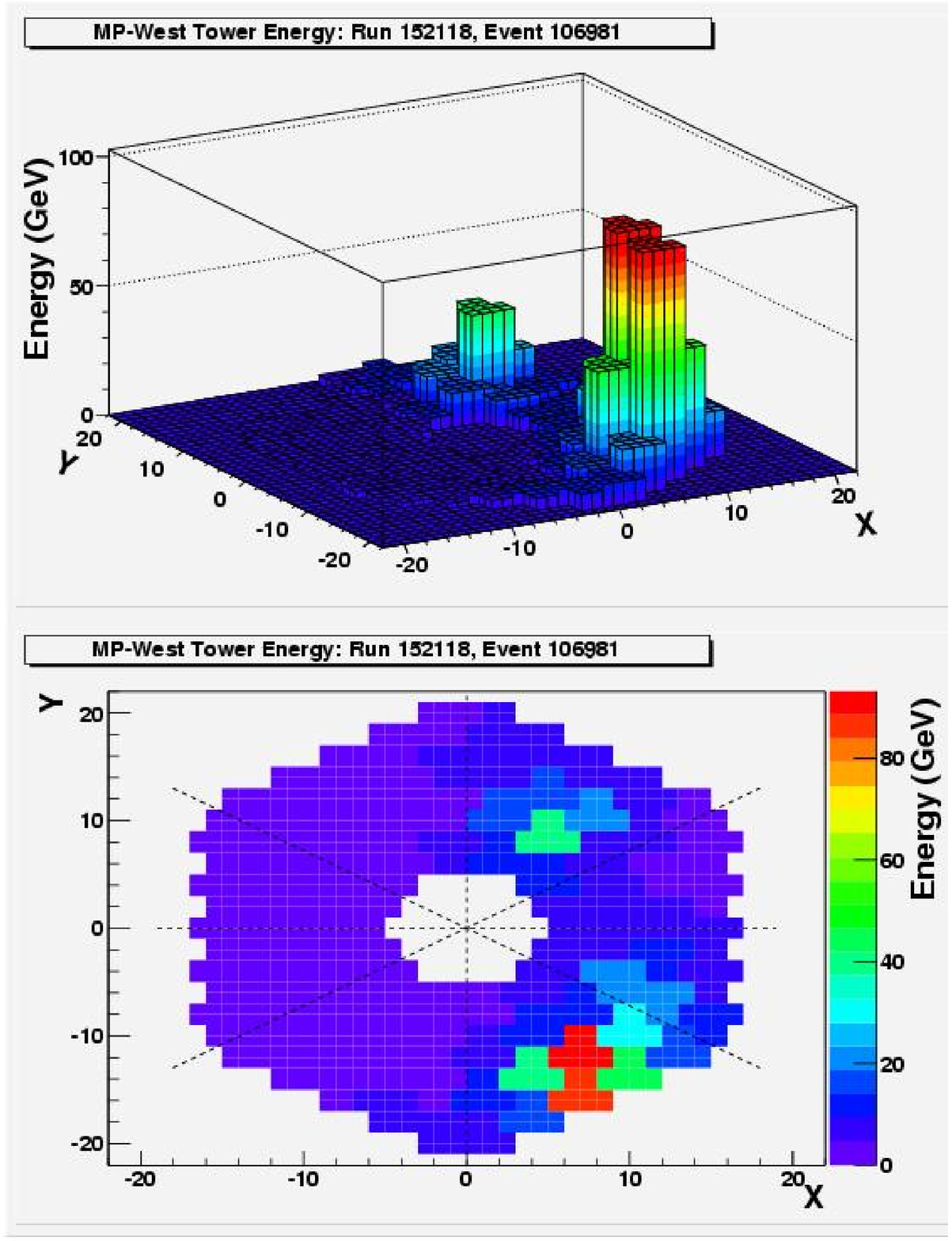}
\vspace*{-1.5cm}
\caption{\label{fig:mp_2jet}
Event display of a two-``jet'' event in the East MP.
The vertical axis shows the signal pulse height
measured in units of GeV.
The term ``jet'' is used to indicate a hadron or
electromagnetic shower and not an actual jet of particles.
The two-dimensional plot (bottom) shows the {\it x-y} 
coordinate of the particles hitting the MP.}
\end{figure}

Figure~\ref{fig:mp_2jet} shows an event display with two jets in the MP
(the term ``jet'' is used here to indicate a cluster of towers, which
is most likely due to one particle interacting in the MP).
This event was selected from a minimum bias data sample.

\section{MEASUREMENT OF $\xi$}

Calorimeter information is used to determine the momentum loss of the anti-proton, 
\begin{equation}
\label{eq:xi_formula}
\hspace*{2cm}\xi_{\overline{p}}=\frac{1}{\sqrt{s}}\sum_{i=1}^nE_T^ie^{-\eta^i}
\end{equation}
which is calculated using the rapidity ($\eta$) and transverse energy ($E_T$) of all calorimeter towers including the MPs.

\begin{figure}[htb]
\includegraphics*[width=\hsize]{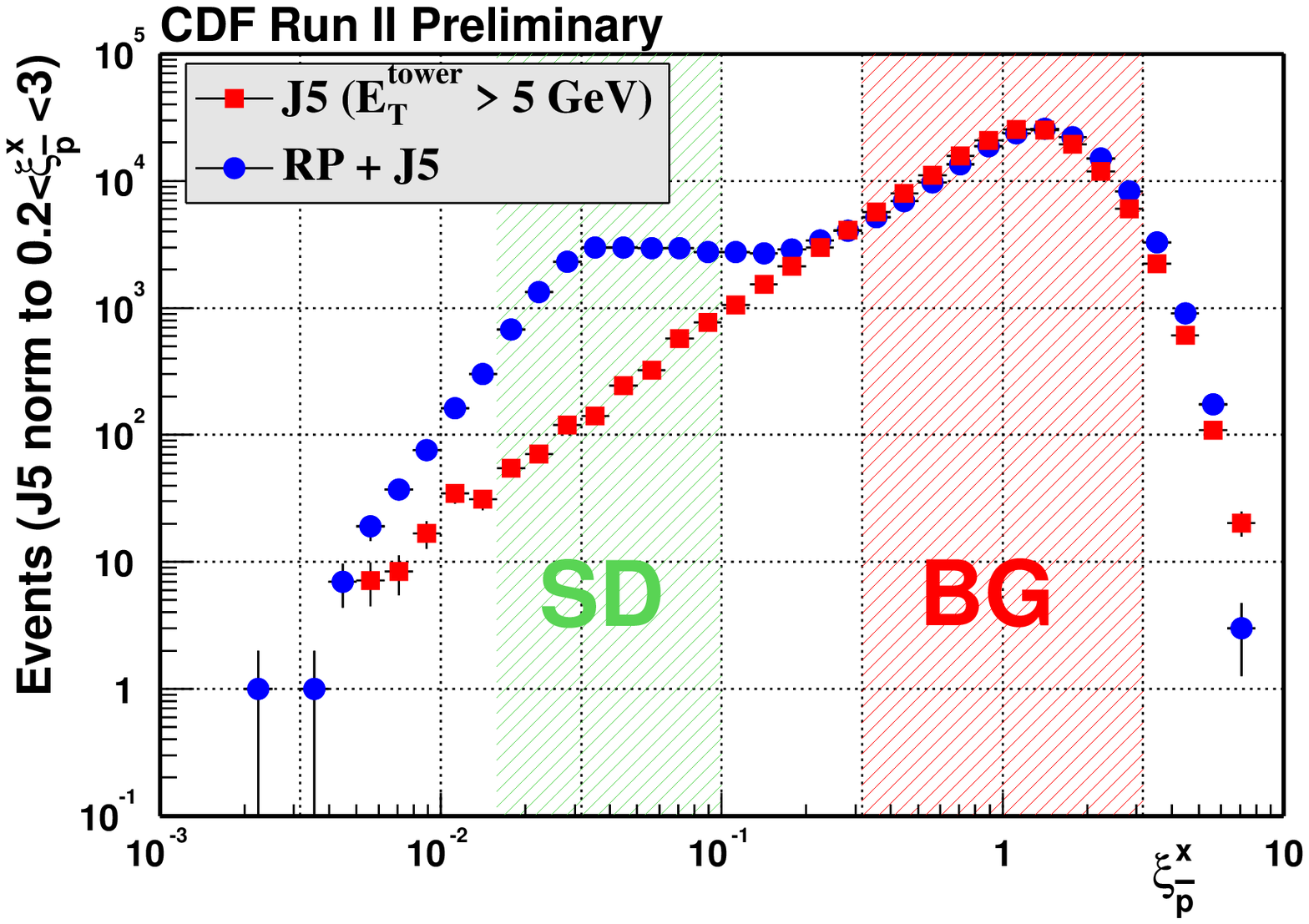}
\vspace*{-1.5cm}
\caption{\label{fig:xi}
Momentum loss of the antiproton ($\xi_{\overline{p}}$) distribution in the RP+J5 and J5 samples.
SD and BG regions are selected according to the measured $\xi$ values.}
\end{figure}

In Run~II, a dedicated trigger (RP+J5) selects events with a three-fold RPS coincidence and at least one 
calorimeter tower with $E_T>5$~GeV.
A further offline selection requires at least two jets of $E_T^{corr}>5$~GeV and $|\eta|<2.5$.
Jet energies are corrected for detector effects and underlying event effects.
A large number of events are at $\xi_{\overline{p}}\sim 1$ (Fig.~\ref{fig:xi}), where the contribution is due to two sources:
diffractive dijets with a superimposed soft non-diffractive interaction, and non-diffractive dijets superimposed with a soft diffractive interaction.
The plateau observed between $0.03<\xi_{\overline{p}}<0.1$ (SD) results from a $d\sigma/d\xi\sim 1/\xi$ distribution, which is expected for diffractive production.
The declining of the distribution below $\xi_{\overline{p}}\sim 0.03$ occurs in the region where the RPS acceptance is decreasing. 
The MP plays an important role, as the contribution to $\xi_{\overline{p}}(\rm CAL)$ from MP towers helps 
separating diffractive from non-diffractive events. The large peak at $\xi_{\overline{p}}\sim 1$ in Figure~\ref{fig:xi} 
would in fact rest on top of the SD region ($\xi_{\overline{p}}\sim 10^{-2}\div 10^{-1}$) 
without the inclusion of the MP towers in the calculation of $\xi_{\overline{p}}$.

\begin{figure}[htb]
\includegraphics*[width=\hsize]{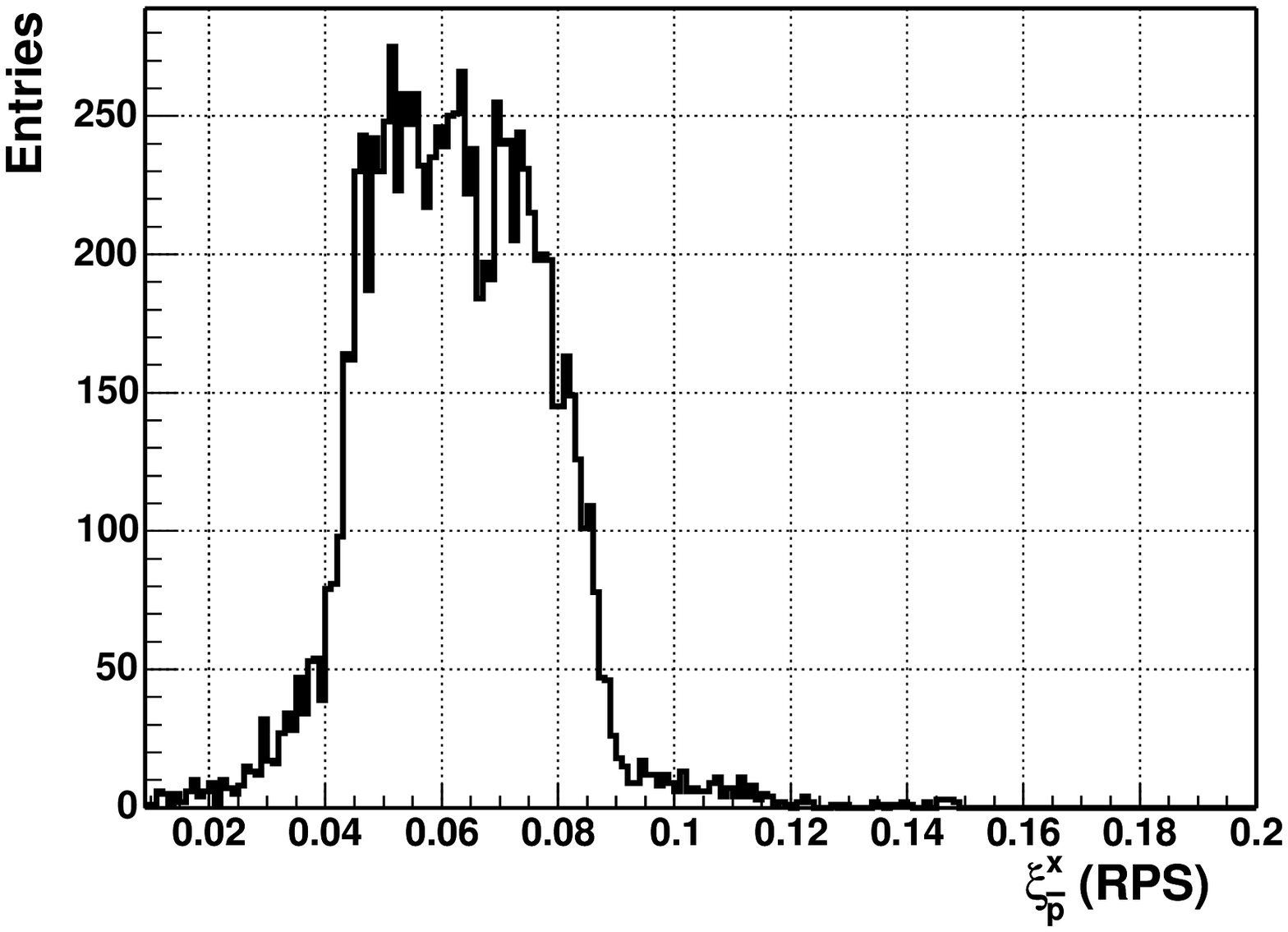}
\vspace*{-1.5cm}
\caption{\label{fig:xi_rp}
Values of antiproton momentum loss as measured in the Roman Pot fiber tracker, $\xi_{\overline{p}}(\rm RPS)$.}
\end{figure}

\begin{figure}[htb]
\includegraphics*[width=\hsize]{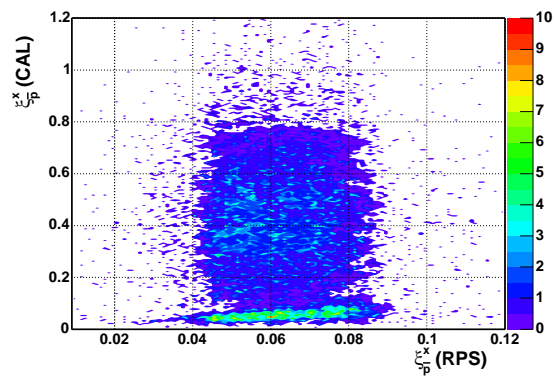}
\vspace*{-1.5cm}
\caption{\label{fig:xi_rp_vs_mp}
Two-dimensional scatter plot of $\xi_{\overline{p}}(\rm CAL)$ versus $\xi_{\overline{p}}(\rm RPS)$ for events with a reconstructed RPS track.}
\end{figure}

The method of measuring $\xi_{\overline{p}}$ using only calorimeter information through Eq.~\ref{eq:xi_formula} is calibrated on the antiproton side by 
comparing the value with that measured by the RPS, $\xi_{\overline{p}}(\rm RPS)$ (Fig.~\ref{fig:xi_rp}).
The two-dimensional scatter plot (Fig.~\ref{fig:xi_rp_vs_mp}) of $\xi_{\overline{p}}(\rm CAL)$ versus $\xi_{\overline{p}}(\rm RPS)$ 
contains data points for tracks reconstructed in the RPS. A linear relationship is observed between $\xi_{\overline{p}}(\rm CAL)$ and $\xi_{\overline{p}}(\rm RPS)$,
in the region where both $\xi_{\overline{p}}(\rm CAL)$ and $\xi_{\overline{p}}(\rm RPS)$ are confined between values of 0.02 and 0.1.
The data points in the region $0.03<\xi_{\overline{p}}(\rm RPS)<0.08$, 
and $\xi_{\overline{p}}(\rm CAL)>0.1$ are presumed to be due to events with an additional superimposed soft non-diffractive interaction. 
Without the calorimeter measurement of $\xi_{\overline{p}}$, these events would represent an inseparable background source to non diffractive events.

\begin{figure}[htb]
\includegraphics*[width=\hsize]{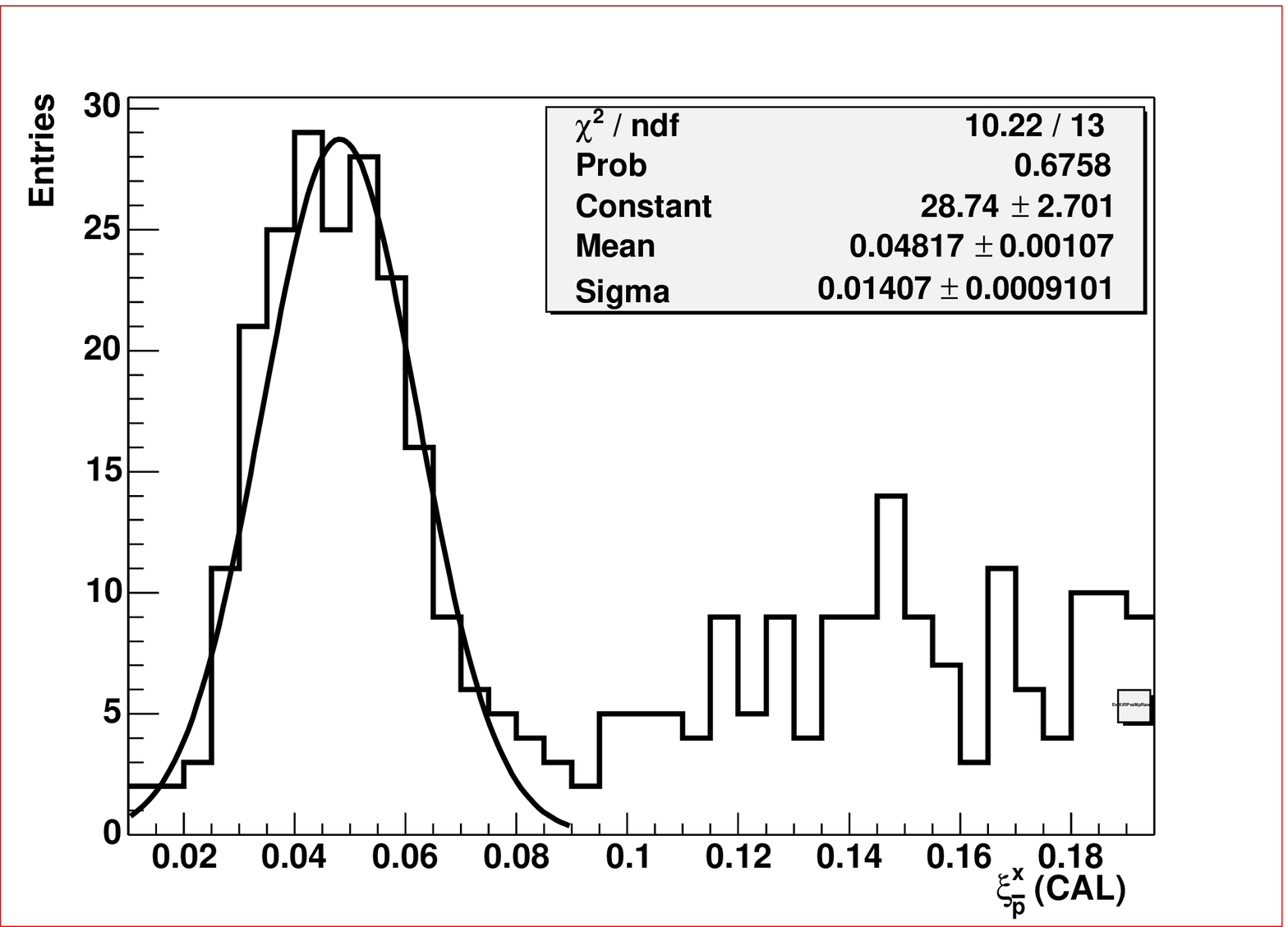}
\vspace*{-1.5cm}
\caption{\label{fig:xi_cal_fit}
Distribution of antiproton fractional momentum loss $\xi_{\overline{p}}(\rm CAL)$, 
measured from calorimeter information for events in which the $\xi_{\overline{p}}(\rm RPS)$ value measured by the Roman Pot fiber tracker is within 
$0.045<\xi_{\overline{p}}(\rm RPS)<0.050$. The line is a gaussian fit.}
\end{figure}

\begin{figure}[htb]
\includegraphics*[width=\hsize]{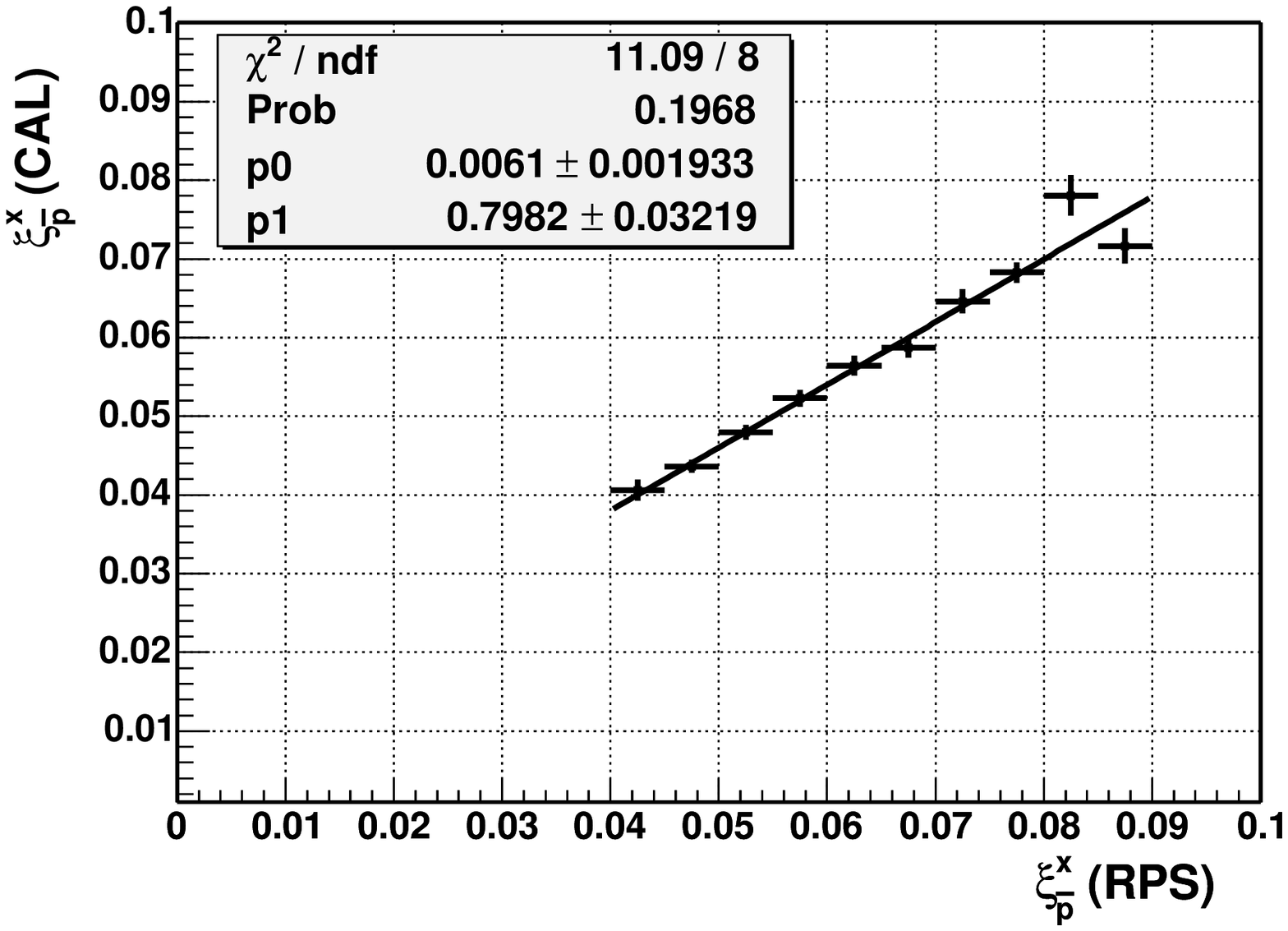}
\vspace*{-1.5cm}
\caption{\label{fig:xi_rp_vs_mp_linfit}
Median values of $\xi_{\overline{p}}(\rm CAL)$ obtained from fits to data in different $\xi_{\overline{p}}(\rm RPS)$ bins, 
plotted versus $\xi_{\overline{p}}(\rm RPS)$. A linear fit describes well the data.}
\end{figure}

For more detailed studies,
data are divided in bins of $\Delta\xi_{\overline{p}}(\rm RPS)=0.005$, and the $\xi_{\overline{p}}(\rm CAL)$ values obtained using calorimeter 
information are fitted with a gaussian distribution for each bin. 
Figure~\ref{fig:xi_cal_fit} shows, as an example, the data and fit for $0.045<\xi_{\overline{p}}(\rm CAL)<0.050$.
The ratio of width to peak position is $\approx 0.3$ over the entire $\xi_{\overline{p}}(\rm CAL)$ region of the data sample.
In the region $0.02<\xi_{\overline{p}}(\rm RPS)<0.1$, an approximately linear relationship is observed (Fig.~\ref{fig:xi_rp_vs_mp_linfit}) between the 
central value of $\xi_{\overline{p}}(\rm CAL)$ and $\xi_{\overline{p}}(\rm RPS)$, $\xi_{\overline{p}}(\rm CAL)=p1\cdot \xi_{\overline{p}}(\rm RPS)$, 
with $p1\simeq 0.80$, which is 20\% lower than the expected value of $p1=1$. 
As the relationship between the two values is linear,
a simple scaling is sufficient to account for the discrepancy between $\xi_{\overline{p}}(\rm CAL)$ and $\xi_{\overline{p}}(\rm RPS)$.

\section{DIFFRACTIVE DIJETS}

Diffractive dijet events are characterized by the presence of two jets resulting 
from a hard scattering and a leading antiproton which escapes the collision intact,
only losing a small momentum fraction $\xi_{\overline{p}}$ to the pomeron.
The gluon and quark content of the interacting partons can be investigated by comparing 
SD and ND events.
The ratio of SD to ND dijet production rates (from Fig.~\ref{fig:xi}) is proportional to the ratio of the
corresponding structure functions and can be studied as a function of the Bjorken
scaling variable $x_{Bj}=\beta\cdot\xi_{\overline{p}}$ of the struck parton in the antiproton, 
where $\beta$ corresponds to the pomeron momentum fraction carried by the parton.
For each event, $x_{Bj}$ is evaluated from the $E_T$ and $\eta$ of the jets using the equation

$$x_{Bj}=\frac{1}{\sqrt{s}}\sum_{i=1}^nE_T^ie^{-\eta^i}$$

In Run~I, CDF measured the ratio of SD to ND dijet production rates using the RPS to detect leading antiprotons. 
The CDF result~\cite{diff_dijet_rp} is suppressed by a factor of $\sim10$ relative 
to predictions from HERA data, indicating a breakdown of
conventional factorization between HERA and the Tevatron.
Correct predictions can be obtained by scaling the rapidity gap probability 
distribution of the diffractive structure function to the total integrated gap probability~\cite{dino2}.

\begin{figure}[htb]
\includegraphics*[width=\hsize]{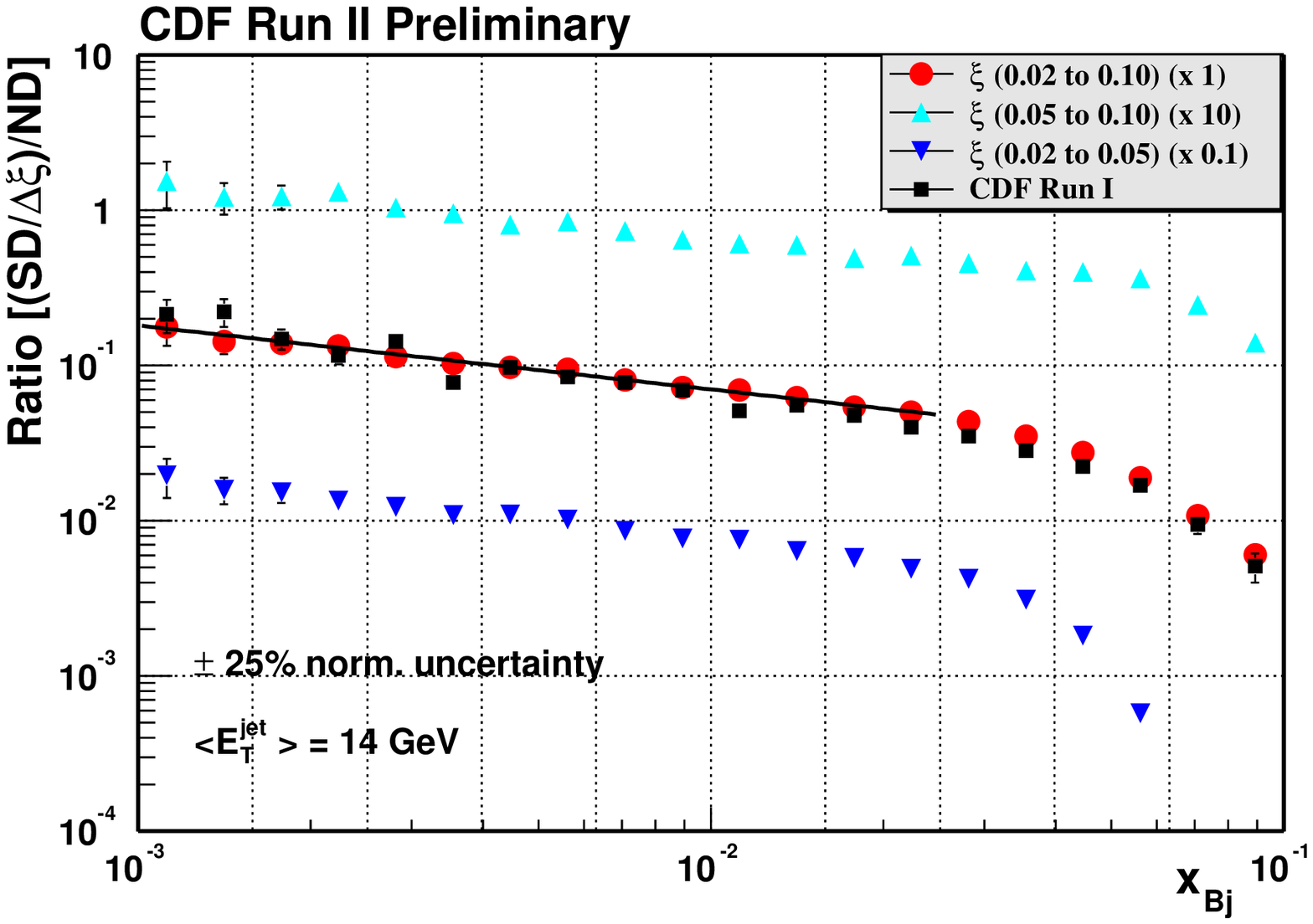}
\vspace*{-1.5cm}
\caption{\label{fig:run2_sf_xi}
Ratio of diffractive to non-diffractive dijet event rates as a function of $x_{Bj}$ compared to Run~I data.}
\end{figure}

\begin{figure}[htb]
\includegraphics*[width=\hsize]{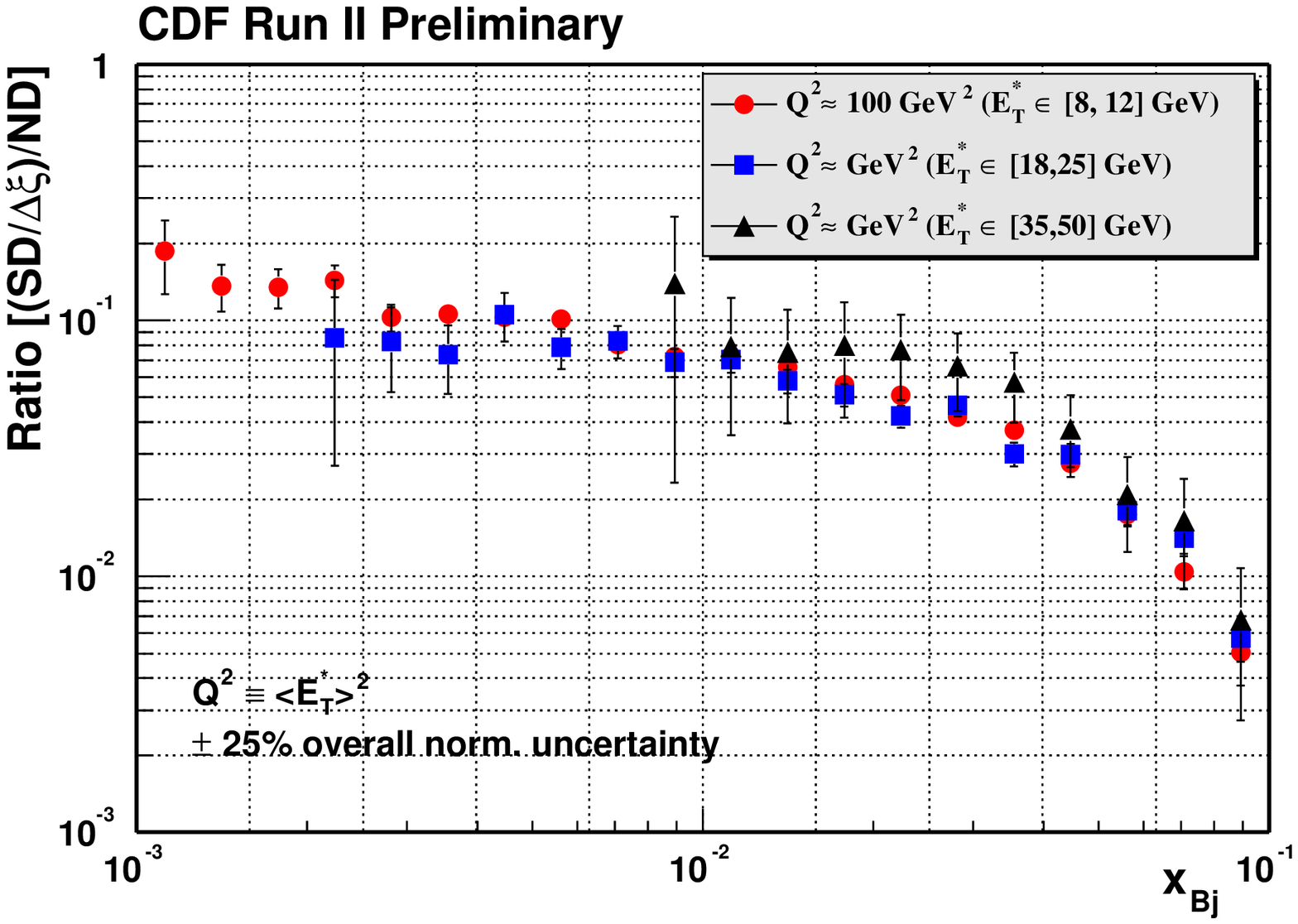}
\vspace*{-1.5cm}
\caption{\label{fig:run2_sf_q2}
Ratio of diffractive to non-diffractive dijet event rates as a function of $x_{Bj}$ for different values of $E{_T}^2 \equiv Q^2 $.}
\end{figure}

Run~II results of a measurement of the SD to ND event rate ratio are consistent with those of Run~I (Fig.~\ref{fig:run2_sf_xi}).
Furthermore, the jet $E_T$ spectrum extends to higher values than in Run~I.
A preliminary result indicates that the ratio does not depend strongly on $E{_T}^2 \equiv Q^2$
in the range from $Q^2 = 100$~GeV$^2$ up to 1600~GeV$^2$ (Fig.~\ref{fig:run2_sf_q2}).
The relative normalization uncertainty cancels out in the ratio.
This result indicates that the $Q^2$ evolution of the pomeron is similar to that of the proton.

\section{CONCLUSIONS}
The program for diffractive physics during Run~II at the Tevatron includes
studies of soft and hard diffraction and of double pomeron exchange.
The Forward Detector upgrade project is an essential component of the diffractive program at the CDF experiment.
It consists of three components: a Roman Pot spectrometer (RPS), two Miniplug (MP) calorimeters, and seven stations of scintillation counters (BSC).
The RPS has been refurbished from the one used in Run~I, and
the BSCs have been added between the IP and the RPS to reject non-diffractive background events. 
The two MP calorimeters were designed to measure the flow of the event energy in the very forward rapidity region.

With the detectors installed, a study of their performance and some of the first results obtained have been discussed here.
In particular, a measurement of the anti-proton momentum loss using calorimeter information (including the MP) was presented.
The results show a linear relationship with the measurement obtained from RPS tracking data.
Results also indicate that RPS tracking data alone are not sufficient to separate diffractive events from non-diffractive background.
The contribution of the MPs to the measurement of $\xi_{\overline{p}}$ allows further discrimination, necessary to reject multiple interactions.
Furthermore, a first measurement performed with the Forward Detectors during Run~II 
re-established the Run~I measurement of the diffractive structure function as the ratio of diffractive to non-diffractive event rates.

The design of the MP calorimeters, capable of working in a high luminosity environment and measuring the energy and lateral position of particles, 
is suitable to be further exploited in the era of the Large Hadron Collider at CERN to extend diffractive physics studies.


\begin{thebibliography}{9}
\bibitem{rapidity}
The terms rapidity and pseudorapidity are used interchangeably.
The pseudorapidity $\eta$ is defined as $-\ln \tan(\theta/2)$, where $\theta$ is the polar angle with respect to the proton beam direction.
The azimuthal angle is denoted by $\phi$, and the transverse energy of a jet, $E_T$, is defined as $E_T\equiv E\cdot \sin\theta$.
\end{thebibliography}
\end{document}